\documentclass[twocolumn,superscriptaddress,aps]{revtex4}
\usepackage{amsmath,amssymb,bm}
\usepackage{graphicx,epstopdf} 
\usepackage{color}
\usepackage{hyperref}

\usepackage{CJK} 

\newcommand{\B}{\mbox{\tiny B}}

\newcommand{\s}{\mbox{\tiny S}}

\newcommand{\wti}{\widetilde}

\newcommand{\nl}{\nonumber \\}

\newcommand{\tL}{\mbox{\tiny L}}
\newcommand{\tR}{\mbox{\tiny R}}
\newcommand{\Sec}[1]{Sec.\;\ref{#1}}

\newcommand{\be}{\begin{equation}}
\newcommand{\ee}{\end{equation}}
\newcommand{\bsube}{\begin{subequations}}
\newcommand{\esube}{\end{subequations}}
\newcommand{\Eq}[1]{Eq.\,(\ref{#1})}

\newcommand{\Fig}[1]{Fig.\,\ref{#1}}

\newcommand{\dg}{\dagger}
\newcommand{\la}{\langle}
\newcommand{\ra}{\rangle}

\begin{document}
 
 
\title{Majorana qubit readout by a point-contact detector under finite bias voltages}

\author{Huizi Xie}
\affiliation{School of Physics, Hangzhou Normal University,
Hangzhou, Zhejiang 311121, China}

\author{Sirui Yu}
\affiliation{School of Physics, Hangzhou Normal University,
Hangzhou, Zhejiang 311121, China}

\author{Hong Mao} \email{mao@hznu.edu.cn}
\affiliation{School of Physics, Hangzhou Normal University,
 Hangzhou, Zhejiang 311121, China}

\author{Jinshuang Jin} \email{jsjin@hznu.edu.cn}
\affiliation{School of Physics, Hangzhou Normal University,
Hangzhou, Zhejiang 311121, China}

\date{\today}

\begin{abstract}

In this work we revisit the problem of a Majorana box qubit (MBQ)
  readout by a point-contact (PC) detector.
The logic states of the MBQ are associated with
 the combined fermion parities 
of the MBQ and its tunnel-coupled quantum dot,
which is measured by a PC detector.
Beyond the existing studies on limiting bias voltage regimes,
we analyze
the steady-state current and the current power spectrum across all bias voltages.
Our results indicate that
 the MBQ readout 
via the parity-dependent detector current 
is effective only at low bias voltage regime
and requires the dot energy level to be off-resonance 
with the Majorana qubit.
In contrast, the current power spectrum allows MBQ 
readout through the parity-dependent
 Rabi oscillation peak signals for arbitrary bias voltages,
without restrictions on
 the dot energy level.
Particularly, with
focus on the MBQ measurement visibility, 
 we analyze the peak-to-pedestal ratio
for each characteristic peak (associated with each logic state of the qubit)
and the signal-to-noise ratio of the two peaks. 
By examining these two metrics,
we identify the optimal bias voltage window
for the PC detector at low temperature limit.

\end{abstract}

\maketitle

\section{Introduction}
 
Owing to the unique nature of non-Abelian statistics and spatial nonlocality,  
Majorana zero modes (MZMs) have a potential application
in topological quantum computations \cite{Kit01131,Kit032,Nay081083,Lei11210502,
Das1515001,Aas16031016,Plu17012001,Kar17235305}. 
Largely based on the MZMs which may be realized 
from the semiconductor/superconductor hybrid structures \cite{Lut10077001,Ore10177002}, 
various types of Majorana qubits have been proposed, 
including the Majorana box qubits (MBQs) \cite{Plu17012001}, tetrons and hexons \cite{Kar17235305}.  
For the Majorana topological quantum computation,
the qubit measurement is an important problem 
to be studied in the relatively early stage.
Indeed, for instance, for the readout of MBQs, 
various schemes have been proposed \cite{ Plu17012001,Gri19235420,Smi20020313,
Qin19043027,Zho21010301,Mun20033254,Ste20033255},
such as measuring the frequency shift of a double-dot system \cite{Plu17012001}, 
dispersive shift from qubit-resonator interaction \cite{Smi20020313},
and interferometric conductance \cite{Qin19043027,Zho21010301}.

\begin{figure}
\includegraphics[width=0.8\columnwidth]{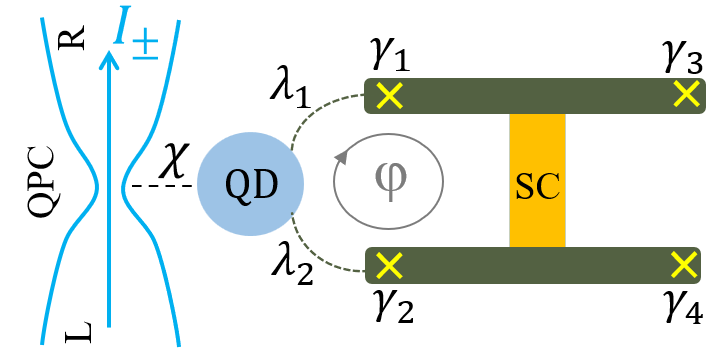} 
\caption{(Color online)
Schematic setup for the readout of  
a Majorana box qubit (MBQ).
The MBQ is coupled to a quantum dot (QD) which is measured
by a point-contact (PC) detector.
The MBQ consists of two
topological superconducting wires, hosting four Majorana
zero energy modes
$\gamma_1,\gamma_2,\gamma_3$ and $\gamma_4$.
The coupling of 
two Majoranas ($\gamma_1$, $\gamma_2$) and the QD leads 
to a parity-dependent
current $I_{\pm}$ through the PC detector. }
\label{fig1}
\end{figure}

In this work, we will revisit the MBQ readout scheme, 
as schematically depicted in \Fig{fig1},
in which a pair of MZMs ($\gamma_1$ and $\gamma_2$) are coupled to a quantum dot (QD) in contact with
 a point-contact (PC) detector \cite{Ste20033255,Mun20033254}.
 The measured current can be classified as $I_{\pm}$, for its being sensitive to 
 the QD-plus-MZM combined parity (${\pm1}$), which presents the two states of MBQ,
 respectively.
   The exiting studies \cite{Ste20033255,Mun20033254} were 
  based on Markovian theory. The main findings are as follows.
  The signatures of parity dependence on the dot occupation number 
  exist only in the low
bias voltage regime \cite{Mun20033254}.
On the other hand, in the large bias voltage regime,
the parity sensitive signals occur in 
the current power spectrum rather than
   average current or dot occupation number   
\cite{Ste20033255}.

The new results of this work include the demonstration of
  the parity-dependent detector current and its power spectrum
across all the bias voltages. 
Particularly, with focus on 
the MBQ measurement visibility,
we obtain the analytical results
and identify the optimal bias voltage window.
To that end, we employ
the Bloch-Redfield master equation (BR-ME) approach 
\cite{Yan982721,Yan002068,Xu02023807}.
Within the BR-ME framework, 
 the transport formulas for the detector current and its power spectrum
have been well-established \cite{Li05066803,Xu22064130}.
Note that the methods exploited in Refs.\cite{Ste20033255,Mun20033254} 
are of the same accuracy as BR-ME. 
Thus, the findings from the earlier studies in limiting bias voltage regimes  
are all reproduced in the present work.

The paper is organized as follows.
 In \Sec{thmeth}, we first give the model description and outline the basic measurement scenario.
 We then introduce the BR-ME to calculate the
 transport current through the detector and the current power spectrum in a unified way.
 In \Sec{thdotI}, we carry out
 the analytical and numerical results of the detector current.
 We also address the underlying physical
 mechanism of the parity-dependent current related to the bias voltage.
In \Sec{thnoise}, the analytical and numerical results for
the current power spectrum are illustrated. Furthermore, we analyze  
 the signal-to-noise ratio (SNR) of the two parity-dependent Rabi peaks and 
  the peak-to-pedestal ratio of each single characteristic peak.
  By examining these two metrics, 
 the optimal bias voltage window for the PC detector is provided.
 Finally, we
give our summary in \Sec{thsum}.

\section{Methodology}
\label{thmeth}

\subsection{Model description}
\label{thmod}

We consider 
a Majorana box qubit (MBQ) coupled to a quantum dot (QD),
with the QD continuously measured
by a voltage-biased point-contact (PC),
 as sketched in
\Fig{fig1}.
The MBQ is formed in a floating superconducting island, which consists of
two long topological superconducting nanowires \cite{Plu17012001,Kar17235305}.
Each nanowire hosts a Majorana zero-energy mode (MZM) denoted as $\gamma_i$ at both ends.
The pairs of MZMs are related to
the regular fermion through the transformations:
 $\gamma_{1}= (  \hat f_{\tL}+  \hat  f^\dg_{\tL})$, $\gamma_{2}= -i(   \hat f_{\tL}-  \hat  f^\dg_{\tL})$
for the left pair, and
$\gamma_{3}= (   \hat f_{\tR}+   \hat f^\dg_{\tR})$, $\gamma_{4}= -i(   \hat f_{\tR}-  \hat  f^\dg_{\tR})$
for the right pair, respectively.
The corresponding occupation number operators are $ \hat n_{\tL}= \hat f^\dg_{\tL} \hat f_{\tL}$
and $ \hat n_{\tR}= \hat f^\dg_{\tR} \hat f_{\tR}$.
The MZMs obey the anticommutation relation $\{\gamma_{i},\gamma_{j}\}=2\delta_{\ij}$
 and satisfy $\gamma^2_{i}=1$.
The qubit states $|n_{\tL}n_{\tR}\ra$ can 
be encoded in the even parity space
$\{|0_{\tL}0_{\tR}\ra,|1_{\tL}1_{\tR}\ra\}$ or
odd parity space $\{|0_{\tL}1_{\tR}\ra,|1_{\tL}0_{\tR}\ra\}$.

The composite Hamiltonian of the device in \Fig{fig1} is
$H_{\rm tot}=H_{\s}+H_{\B}+H'$.
The system is modeled by the low-energy Hamiltonian \cite{Mun20033254,Fle11090503},
 \begin{equation}\label{Hs}
  H_{\s}=\varepsilon_{\rm d}  \hat d^\dg  \hat d  
 +\sum_{i=1,2}\big(\lambda_{i}\gamma_{i} \hat  d
 +{\rm H.c.}\big).
 \end{equation}
Here, we assume 
a pair of MZMs described by $\gamma_{1}$ and $\gamma_{2}$
and a single electronic level $\varepsilon_{\rm d}$ in the QD.
The second term in \Eq{Hs} describes the tunnel coupling between the QD and two MZMs.
Without loss of generality, the tunneling coefficients $\lambda_i$
can be defined as,
 $\lambda_{1}=|\lambda_1|e^{i\varphi_1}$ and $\lambda_{2}=|\lambda_2|e^{i\varphi_2}$,
with the gauge invariant phase difference $\varphi=\varphi_2-\varphi_1$ tunable,
for instance,
 by varying the magnetic flux penetrating the enclosed area
of the interference loop (see \Fig{fig1}).
The PC detector is represented by the two electron reservoirs, described by the Hamiltonian
$H_{\B}=\sum_{\alpha k}\varepsilon_{\alpha k} \hat c^\dg_{\alpha k} \hat c_{\alpha k}$
with $\alpha={\rm L,R}$.
The interaction Hamiltonian between the dot and the PC detector
is given by \cite{Gur9715215,Kor995737,Goa01235307}
\begin{align}\label{Hint}
H'&=\sum_{k,q}({\cal T}+\chi  n_{\rm d}) \hat c^\dg_{{ \tL}k} \hat c_{{\tR}q}+{\rm H.c.},
\end{align}
where ${\cal T}$ represents the tunnel coupling amplitude independent of the 
QD occupation, and $\chi$ is small parameter indicating weak interaction with 
$\chi\ll{\cal T}$.
For convenience,
we introduce the system-related operator
$ \hat Q={\cal T}+\chi  \hat n_{\rm d}= \hat Q^\dg$ and the PC bath operator
$ \hat F=\sum_{kq} \hat c^\dg_{\tL k} \hat c_{{\tR}q}$.
The interaction Hamiltonian \Eq{Hint}
is then straightforwardly rewritten as $H'= \hat Q( \hat F+ \hat F^\dg)$.

The measurement effects of PC reservoirs on
the combined QD-MBQ system are characterized by the interaction
bath correlation functions \cite{Li04085315,Li05066803}:
\be\label{ct}
C^{(+)}(t)\equiv\la  \hat F^\dg(t)  \hat F(0) \ra_{\B},\,\, C^{(-)}(t)\equiv\la  \hat F(t)  \hat F^\dg(0) \ra_{\B},
\ee
with $ \hat F(t)=e^{iH_{\B}t} \hat Fe^{-iH_{\B}t}=\sum_{kq} \hat c^\dg_{\tL k} \hat c_{{\tR}q}
e^{i(\varepsilon_{\tL k}-\varepsilon_{\tR q})t}$.
Here, $\la \cdots \ra_{\B}$ stands for the statistical average over the bath (PC electron reservoirs)
in thermal equilibrium.
Throughout this work, we adopt units of $e=\hbar=1$
for the electron charge and the Planck constant.

\subsection{Basic measurement scenario}
\label{thmea}

Taking the even parity space ($\{|0_{\tL}0_{\tR}\ra,|1_{\tL}1_{\tR}\ra\}$) as a typical example,
we provide a brief overview of the basic measurement protocol for the MBQ
\cite{Mun20033254,Ste20033255}.
Throughout the measurement process, 
 the total parity of the superconducting island is assumed to be conserved.  
Therefore,
the readout of the MBQ states $|0_{\tL}0_{\tR}\ra$ and $|1_{\tL}1_{\tR}\ra$
can map to the readout of the states 
$|0_{\tL}\ra$ and $|1_{\tL}\ra$, respectively.
Then we can denote $|0_{\tL}\ra$ ($\rightarrow|0_{\tL}0_{\tR}\ra$) and 
$|1_{\tL}\ra$ ($\rightarrow|1_{\tL}1_{\tR}\ra$) as a logical qubit states.
The readout of this logical qubit states can be realized by the measurement of the combined fermion parity 
  of the dot and the left pair of MZMs ($\gamma_1$ and $\gamma_2$), defined as
\begin{equation}\label{parity}
p=(-1)^{n_{\rm d}+n_{\tL}},
\end{equation}
since this parity is conserved by
the total composite Hamiltonian $H_{\rm tot}$.
For instance,
the MBQ is assumed to be prepared in the initial state
e.g., $|\psi_0\ra=a|0_{\tL}\ra+b|1_{\tL}\ra$,
with the complex coefficients $a$ and $b$ satisfying $|a|^2+|b|^2=1$,
and the dot is supposed initially empty, $n_{\rm d}=0$. 
The initial state of the dot-MBQ can be expressed as:
 \begin{equation}\label{psi0}
|\Psi_0\ra=|0_{\rm d}\ra\otimes(a|0_{\tL}\ra+b|1_{\tL}\ra)
=a|0_{\rm d}0_{\tL}\ra+b|0_{\rm d}1_{\tL}\ra.
\end{equation}
Thus, the MBQ state can be determined: if
$p=+1$, the state is $ |0_{\tL}\ra$, whereas if  
$p=-1$, the state is $ |1_{\tL}\ra$.

The dot-MZM interaction Hamiltonian of \Eq{Hs} entails
intrinsic coherent Rabi oscillations 
within both  
the odd parity subspace $\{ |0_{\rm d}1_{\tL}\ra, |1_{\rm d}0_{\tL}\ra \}$ (${ p}=-1$)
and even parity subspaces $\{|0_{\rm d}0_{\tL}\ra, |1_{\rm d}1_{\tL}\ra\}$ (${ p}=+1$).
The former is induced by the electron normal tunneling processes, whereas
the latter arises from the Andreev reflection processes.
Consequently, the Hamiltonian \Eq{Hs} can be expressed as,
\begin{equation}\label{Hsm}
H_{\s}={\left(\begin{array}{cc}
           H_{-} & 0 \\
           0 & H_{+}
         \end{array}
         \right)},
\end{equation}
where the parity-dependent Hamiltonians in each
  subspace are given by
\begin{equation}\label{Hpm}
H_{ {\rm p}=\pm}={
  \left( \begin{array}{cc}
         0 & \lambda_{1}-i{ p} \lambda_{2}\\
         \lambda^*_{1}+i{ p} \lambda^*_{2} & \varepsilon_{\rm d}
       \end{array}
       \right)}.
\end{equation}
Evidently, $H_{\rm p}$ indicates
 the intrinsic coherent Rabi oscillation.
 The corresponding Rabi frequency 
($\Delta_{\rm p}$) reads, 
 \be\label{deltap0}
\Delta_{\rm p}=\sqrt{\varepsilon^2_{\rm d}
  +4\lambda^2_{\rm p}(\varphi)},
\ee
where $$\lambda^2_{\rm p}(\varphi)\equiv|\lambda_1|^2+|\lambda_2|^2+2p|\lambda_1||\lambda_2|\sin\varphi.$$
Note that the Rabi frequency $\Delta_{\rm p}$ 
  is the energy difference between the two eigenstates
  ($\{|e_{\rm p}\ra,|g_{\rm p}\ra\}$) of
$H_{\rm p}$,
i.e., $\Delta_{\rm p}= \varepsilon_{e_{\rm p}}- \varepsilon_{g_{\rm p}}$,
 with the excited energy
$\varepsilon_{e_{\rm p}}=(\varepsilon_{\rm d} +\Delta_{\rm p})/2$
and the ground energy $\varepsilon_{ g_{\rm p}}=(\varepsilon_{\rm d} -\Delta_{\rm p})/2$.
 The resulting Rabi frequency expressed in \Eq{deltap0} is 
a periodic function of $\varphi$ with $2\pi$ periodicity
and exhibits parity-dependent for $\varphi\neq k\pi$.
Consequently, the MBQ can, in principle, be readout  
through any observable quantities that depend on $\Delta_{\rm p}$
when $\varphi\neq k\pi$, including
  the dot occupation number \cite{Mun20033254}, 
  the detector current, and the current power spectrum \cite{Ste20033255}.

\subsection{Quantum master equation approach}
\label{thME}
We now briefly introduce the Bloch–Redfield master equation approach \cite{Yan982721,Yan002068,Xu02023807},
which has been generalized to incorporate the transport current
 and its power spectrum
\cite{Li05066803,Xu22064130}.  
As is well-known, the primary intrested QD-MBQ system is described by
 the reduced density operator
 $\rho(t)\equiv{\rm tr}_{\B}[\rho_{\rm tot }(t)]$,
i.e., the partial trace of the total density operator
$\rho_{\rm tot}$ over the bath space
(the electron reservoirs of PC detector).
By treating the interaction Hamiltonian $H'$ in \Eq{Hint} 
as a perturbation
 and adopting the standard Born-Markovian approximation,
one can easily get the Bloch–Redfield master equation for QD-MBQ system.
It reads \cite{Yan982721,Yan002068,Xu02023807},
 \begin{align}\label{QME}
 \dot\rho(t)&=-i{\cal L}\rho(t)-{\cal R}\rho(t),
\end{align}
where the first term, ${\cal L}\bullet \equiv[H_{\s},\bullet]$, describes the
coherent dynamics and the second term,
\begin{align}
{\cal R}\bullet\equiv \frac{1}{2}\big[Q, {\cal Q} \bullet-\bullet{\cal Q}^\dg \big],
\end{align}
arises from the measurement back-action of the PC detector.
The resolvant operators are $Q={\cal T}+\chi n_{\rm d}$ and
 ${\cal Q} = {\cal Q}^{(-)} +{\cal Q}^{(+)} $,
with ${\cal Q}^{(\pm)} \equiv   C^{(\pm)}(-{\cal L}) Q$.
The bath correlation function $C^{(\pm)}(-{\cal L})$  
is the spectrum 
of the reservoir-electron correlation function \Eq{ct}, 
%
 \begin{align}\label{cw}
C^{(\pm)}(-{\cal L})&= \int^\infty_{-\infty}{\rm d}t\, e^{-i{\cal L} t} C^{(\pm)}(t)
 =\frac{\eta x}{1-e^{-\beta x}} \Big|_{x=-{\cal L}\mp V},
\end{align}
which satisfies the detailed-balance relation: $C^{(+)}(\omega)
=e^{\beta(\omega-V)}C^{(-)}(-\omega)$ for arbitrary bias voltage $V=\mu_{\rm L}-\mu_{\rm R}$.
Here, we introduced the parameter
$\eta=2\pi g_{\rm L}g_{\rm R}$
considering the energy-independent density of
states ($g_{\rm L}$
and $g_{\rm R}$  of the two reservoirs).

The Liouvillian operator``${\cal L}$" in the correlation
function $C^{(\pm)}(-{\cal L})$ contains the
information about energy transfer between the detector and
the measured system.
In the eigen-state basis $\{|e_{\rm p}\ra,|g_{\rm p}\ra\}$, ``${\cal L}$"
is replaced by ``$\Delta_{\rm p}$" which enters the correlation function
as $C^{(\pm)}(\pm\Delta_{\rm p})$.
All the observable quantities measured by PC detector, as demonstrated below,
are closely related to $C^{(\pm)}(\pm\Delta_{\rm p})$,
 thus carrying the parity information.

For the calculation of the current through PC detector and
its power spectrum, one can extend the master equation of \Eq{QME}
into the number-resolved (conditional) master equation
and/or its conjugate counting
field formalism \cite{Zwa01,Mak01357}.
Alternatively, one can introduce
the current-related
density operator based on
 the energy-dispersed dissipaton
decomposition \cite{Xu22064130}.
Within the framework of the Bloch-Redfield theory, 
we can derive the equivalent expressions for the detector current 
and its power spectrum using the aforementioned two methods.
Explicitly, the average current through PC detector 
is given by \cite{Li05066803,Xu22064130}
\begin{align}\label{currt}
I(t)&= {\rm tr}
\big[{\cal J}^{(-)}\rho(t) \big],
\end{align}
and its power spectrum
can be calculated as
\begin{align}\label{noiseform}
 S(\omega)&\!=\!
 2  {\rm tr}\!\big[{\mathcal {J}}^{(+)}\bar\rho\big]
   \!+\!4  {\rm tr}\!\big[{\mathcal {J}}^{(-)}
   {\Pi}(\omega){\cal J}^{(-)} \bar\rho\big].
\end{align}
   Here, the superoperators are defined as
   \bsube
      \begin{align}
 {\mathcal {J}}^{(\pm)}  \bullet&\equiv\frac{1}{2}\big[{\cal Q}^{(-)}\pm{\cal Q}^{(+)}\big]\bullet Q+{\rm H.c.},
  \\
  \label{piw}
   {\Pi}(\omega)&\equiv\frac{1}{i({\cal L}-\omega)+{\cal R}},
 \end{align}
 \esube
 and the stationary state $\bar\rho\equiv\rho(t\rightarrow\infty)$ is obtained by setting $\dot{\rho}(t)=0$
 in \Eq{QME}.

Note that 
 in the following numerical calculations,
 all energies are measured in (arbitrary)
units of $\lambda$.
 We set $|\lambda_1|=\lambda$ and $|\lambda_2|=0.8 \lambda$
 for the consideration of
a certain degree of the coupling asymmetry between the QD and MZMs 
as observed in experiments.
Furthermore, we consider a weak interaction between the PC detector and the QD
with ${\cal T}= \lambda$ and $\chi=0.2 \lambda$ as stated in \Eq{Hint}.
Throughout the paper, we adopt
the energy-independent density of states with $g_{\tL}=g_{\tR}= \lambda$
 and finite temperature $k_{\rm B}T=0.5 \lambda$, unless otherwise stated.

 \begin{figure}
\centering
\includegraphics[width=0.5\textwidth,clip=true]{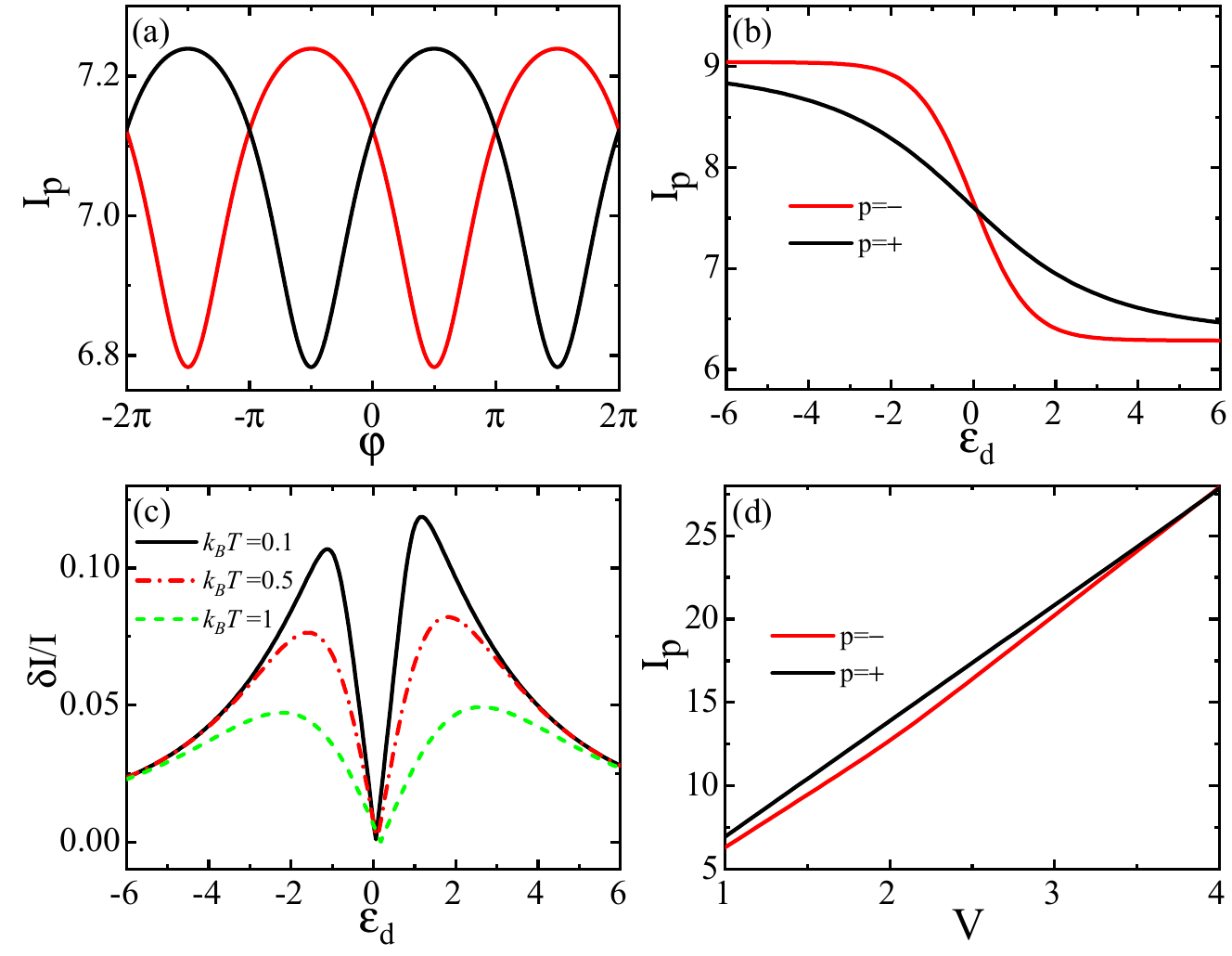}
 \caption{ (Color online) The parity-dependent steady-state current $I_{\rm p}$ (in $e\lambda/\hbar$) 
 at low bias voltage $V=\lambda$,
 (a) as a function of flux $\varphi$
 with $\varepsilon_{\rm d}=\lambda$ and (b) as a function of $\varepsilon_{\rm d}$ 
 with $\varphi=\pi/2$.
  (c) The signal-to-current ratio as a function of $\varepsilon_{\rm d}$ with different temperature (in $\lambda$)
 for $\varphi=\pi/2$ and $V=\lambda$.
 (d) The detector current $I_{\rm p}$ as a function of the bias voltage for 
 $\varphi=\pi/2$ and $\varepsilon_{\rm d}=2\lambda$.
   \label{fig2}}
\end{figure}

\section{The steady-state current}
\label{thdotI}
We first analyze the characteristics of the steady-state current through the PC detector.
By substituting $\rho(t)$ with $\bar\rho$ in the expression of \Eq{currt},
we obtain the steady-state current as
\begin{align}\label{currp}
 I_{\rm p}
&=g_u V- \frac{1}{2} (g_u-g_v)V\Big[1+  \frac{\Delta_{\rm p} }{G^{(+)}(\Delta_{\rm p})}\Big]
\nl&\quad
+g_b V\big[1-\frac{\Delta_{\rm p}}{V}
\frac{G^{(-)}(\Delta_{\rm p})}{G^{(+)}(\Delta_{\rm p})}\big],
 \end{align}
 where $g_u=\eta u^2 $, $g_v=\eta v^2 $, and $g_b=\eta|b|^2$,
  with $u= {\cal T}+\frac{\chi}{2}(1+\frac{\varepsilon_{\rm d}}{\Delta_{\rm p}})$,
$v={\cal T}+\frac{\chi}{2}(1-\frac{\varepsilon_{\rm d}}{\Delta_{\rm p}})$,
and $|b|^2= \frac{\chi^2\lambda^2_{\rm p}(\varphi)}{\Delta^2_{\rm p}}
  $.
Here, we introduce
\bsube\label{Gpm}
\begin{align}
 G^{(+)}(\Delta_{\rm p})&\equiv \frac{C(\Delta_{\rm p})+C(-\Delta_{\rm p})}{2\eta},
 \label{GP}
 \\
  G^{(-)}(\Delta_{\rm p})&\equiv \frac{\bar C(\Delta_{\rm p})-\bar C(-\Delta_{\rm p})}{2\eta},
  \end{align}
  \esube
  with
  $  C(\omega)\equiv C^{(-)}(\omega)+C^{(+)}(\omega)$ and
 $\bar{C}(\omega)\equiv C^{(-)}(\omega)-C^{(+)}(\omega)$.
The analytical expression for the steady-state current $I_{\rm p}$ 
provided in \Eq{currp}
is one of the main results of this work.
We observe that the detector stationary current $I_{\rm p}$  
 depends on $\Delta_{\rm p}$
 through the bath correlation function $C^{(\pm)}(\pm \Delta_{\rm p})$
 and its periodic variation with the flux 
 is illustrated in \Fig{fig2} (a).
 Thus, the MBQ can be effectively readout from $I_{\rm p}$ for $\varphi\neq k\pi$.
 Notably, the signal $\delta I=|I_+-I_-|$ reaches its maximum at $\varphi=(2k+1)\frac{\pi}{2}$,
where $k$ is an integer.

The parity-dependent average current ($I_{\rm p}$)
exhibits a similarity to the
steady-state dot occupation number ($n_{\rm p}$).
From the definition $n_{\rm p}=\la \hat d^\dg \hat d\ra_{\rm p}$,
one can readily recover the result
 $n_{\rm p}  =\frac{1}{2}
\big[1- \frac{ \varepsilon_{\rm d} }{G^{(+)}(\Delta_{\rm p}) }\big]$
 as reported in Ref.\,\onlinecite{Mun20033254}.
The dot occupation number $n_{\rm p}$ is also parity-dependent  
 through the bath correlation function of $G^{(+)}(\Delta_{\rm p})$ (given by \Eq{GP}),
 thus allowing for the readout of the MBQ.
Note that 
for $\varepsilon_{\rm d}=0$, 
the dot occupation number is simplified to $n_{\rm p}=\frac{1}{2}$ 
and the steady-state current becomes parity-independent 
 as displayed in \Fig{fig2} (b).
   This occurs because when $\varepsilon_{\rm d}=0$,
  the coherent dynamics between the two states in each subspace, i.e., $|00\ra$ ($|01\ra$) and $|11\ra$ ($|10\ra$) 
  for even (odd) parity, are equivalent Rabi resonances. 
Furthermore, we plot the signal-to-current ratio, i.e., $\delta I/I=|I_+-I_-|/I$ with $I=\frac{I_++I_-}{2}$,
 as depicted in \Fig{fig2} (c).
 It indicates that the readout of MBQ from
 the detector current $I_{\rm p}$ should away from (but close to)
 the resonance at $\varepsilon_{\rm d}=0$ in the low temperature regime
 with $k_BT\ll \lambda$. 

Figure \ref{fig2} (d) displays
 the steady-state current 
as a function of the bias voltage.
 Evidently,  %
\emph{in the large bias voltage regime}, i.e., $V\gg \Delta_{\rm p}$,
the current becomes parity-independent and thus cannot be used to readout of the MBQ.
The behind reason can be understood as follows.
According to the expression in \Eq{currp},
the involved correlation functions simplify to be energy-independent, i.e.,
$C^{(+)}(\pm\Delta_{\rm p})=0$ and $C^{(-)}(\pm\Delta_{\rm p})=\eta(V\pm\Delta_{\rm p})
\rightarrow \eta V$ as inferred from \Eq{cw}
under large bias voltage limit.
Physically, this is because the energy emitted/absorpted by the system ($\Delta_{\rm p}$)
is negligible compared to the large bias voltage.
As a result,
the average current in \Eq{currp} converges to
 \begin{align}\label{currLV}
 I_{\rm p}\rightarrow I_{\rm c}
&=g_0 V+g_1 V,
 \end{align}
 which is independent of the parity.
 Here, $g_0=\eta({\cal T}+\chi/2)^2$ and $g_1=\eta(\chi/2)^2$.
 %
  %
 In fact, \Eq{currLV} can be easily obtained as
$I_{\rm c}=(I_1+I_0)/2$ with $I_1=\eta
({\cal T}+\chi)^2V$ and $I_0=\eta {\cal T}^2V$, which 
correspond to the currents for the dot states $|1\ra$ and $|0\ra$,
respectively. 
Consequently, it becomes impossible to readout of
the MBQ from the detector current or the steady-state dot occupation number ($n_{\rm p}\rightarrow1/2$)
in the large bias voltage regime. 
 This finding is consistent with the results demonstrated 
 in Refs\,\cite{Mun20033254,Ste20033255}.

Thus, regarding the steady-state current and the dot occupation number, 
 two conditions must be met for the experimental readout of the parity number.
Firstly, the energy level of the dot should be tuned away from the resonance ($\varepsilon_{\rm d}\neq 0$).
 Secondly, the applied bias voltage 
 should be low ($V\lesssim\Delta_{\rm p}$).

\section{Power spectrum}

\label{thnoise}

As is well-known, the current power spectrum contains the information
about the system that goes beyond the average current. From
the formula \Eq{noiseform},
it can be inferred that
the current power spectrum characterizes
  not only the back action of the detector described by the first term, 
but also the intrinsic coherent dynamics
of the measured system depicted by the second term.
The latter directly correlates the current power spectrum 
 with the
parity-dependent Rabi frequency $\Delta_{\rm p}$.
Consequently, the current power spectrum is parity-dependent 
without being limited by the dot energy level
and/or the bias voltage.
Specifically, as long as  $\delta \Delta\equiv |\Delta_{+}-\Delta_-|\neq 0$ 
with $\varphi\neq k\pi$, the MBQ can be extracted from the
current power spectrum, as detailed below.
For clarity, we focus on the resonance case where $\varepsilon_{\rm d}=0$ to 
derive an analytical result for the power spectrum.
Additionally, we adopt $\varphi= \pi/8$ as a typical example
for subsequent numerical calculations.

    \begin{figure}
\centering
\includegraphics[width=0.45\textwidth,clip=true]{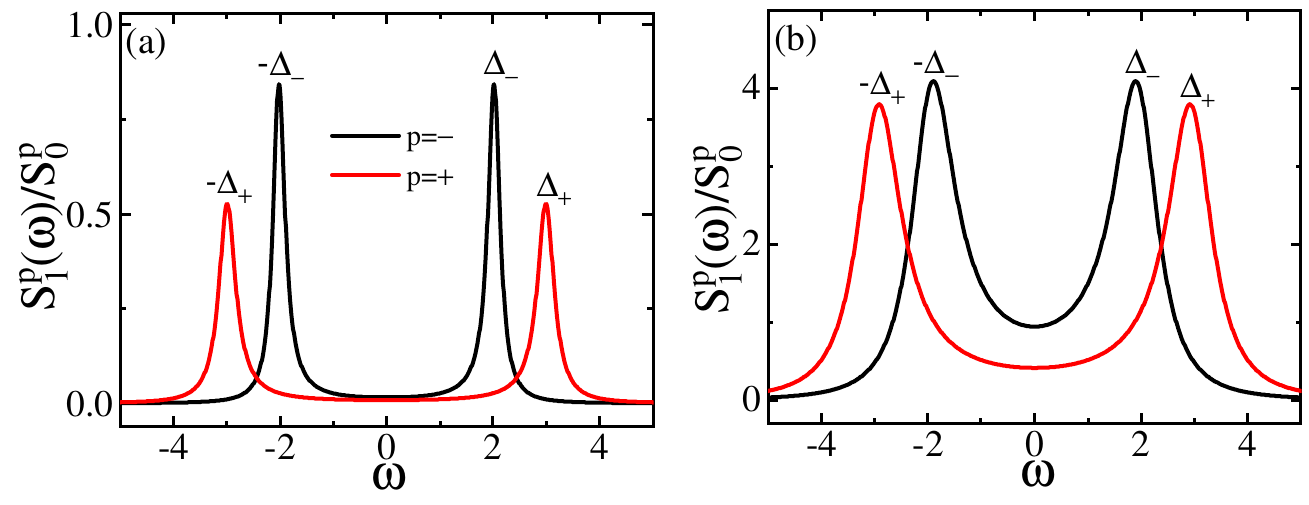}
 \caption{ (Color online) The finite-frequency current power spectrum for $\varepsilon_{\rm d}=0$
 and $\varphi= \pi/8$ with 
 (a) at low bias voltage $V=\lambda$  
 and (b) at large bias voltage $V=8\lambda$.
  \label{fig3}}
\end{figure}

Based on \Eq{noiseform},
we derive the parity-dependent power spectrum denoted by 
$S_{\rm p}(\omega)=S^{\rm p}_0+S^{\rm p}_1(\omega)$, where
 $S^{\rm p}_0$  is a frequency-independent part and $S^{\rm p}_1(\omega)$
is a frequency-dependent part.
 Explicitly, we obtain:
 \bsube\label{SwEd0}
\begin{align}
\label{Sp0}
  S^{\rm p}_0 &= 2g_0 V \coth\frac{V}{2T} + 2g_1
               \Big[ G^{(+)}_{\rm p}-\frac{\Delta^2_{\rm p}}{G^{(+)}_{\rm p}} \Big] ,
  \\
S^{\rm p}_1(\omega)
\label{Sp1wEd0}
&=
       \frac{F_1(\Delta_{\rm p})\Gamma_{\rm p}\Delta^2_{\rm p}}
       {(\omega^2-\Delta^2_{\rm p})^2 +\Gamma_{\rm p}^2\omega^2}
       +
    \frac{F_2(\Delta_{\rm p})G^{(-)}_{\rm p}}{\omega^2+\Gamma_{\rm p}^2},
\end{align}
\esube
where $G^{(\pm)}_{\rm p}\equiv G^{(\pm)}(\Delta_{\rm p})$, $\Gamma_{\rm p}=2g_1G^{(+)}_{\rm p}(\Delta_{\rm p})$,
$F_1(\Delta_{\rm p})=I^2_{\rm d}\Big[1\!-\!\frac{\Delta_{\rm p}}{2V}\frac{G^{(-)}_{\rm p}}
       {G^{(+)}_{\rm p}} \Big]$ with $I_{\rm d}\!=\!I_1\!-\!I_0$,
       $F_2(\Delta_{\rm p})= 8g^3_1 \big[2\Delta_{\rm p}V
             -\Delta^2_{\rm p}\frac{G^{(-)}_{\rm p}}{G^{(+)}_{\rm p}}
             - G^{(-)}_{\rm p}G^{(+)}_{\rm p} \big]$. 
The corresponding numerical results are illustrated in \Fig{fig3} (a)
at low bias voltage ($V=\lambda$)
 and \Fig{fig3} (b) at large bias voltage ($V=8\lambda$), respectively.
Evidently, the MBQ can be read out from the 
 locations of the parity-dependent peaks around the Rabi frequencies, 
 $\omega\simeq\pm\Delta_{\rm p}$, in the power spectrum at all bias voltage.
The current power spectrum of \Eq{SwEd0}, which constitutes the second major result
of this work,
generalizes the result in Ref.\onlinecite{Ste20033255} for the high bias voltage limit
  to the arbitrary bias voltages.

From \Fig{fig3},
we observe that increasing  
 the measurement voltage yields two effects on the 
power spectrum: an augmentation in the peak heights and 
an enhancement of the peak widths.
The former is advantageous for the measurement, while the latter is disadvantageous.
Thus, it is essential to determine the optimal bias voltage range 
of the PC detector for the visibility of qubit measurements.

 Firstly, to characterize the effect of voltage on peak width, 
 we recast the 
frequency-dependent power spectrum formula \Eq{Sp1wEd0}
into the Lorentz form. After some algorithms, we obtain
\begin{align}\label{Sw-Loren}
S^{\rm p}_1(\omega)&\simeq 
\frac{4 f(\Delta_{\rm p})F_1(\Delta_{\rm p})\Delta_{\rm p}^2  }
{\Gamma_{\rm p}(4\Delta_{\rm p}^2-\Gamma_{\rm p}^2) }
\sum_{\pm}\frac{ \sigma^2_{\rm p}}{ (\omega\pm\wti\Delta_{\rm p})^2+ \sigma^2_{\rm p}}
\nl&\quad
+\frac{g(\!\Delta_{\rm p}\!)F_1(\!\Delta_{\rm p}\!)\Gamma_{\rm p} }{\Delta_{\rm p}^2}
\frac{ \sigma^2_{0\rm p}}{ \omega^2\!+\! \sigma^2_{0\rm p}}
\!+\!\frac{F_2(\Delta_{\rm p})G^{(-)}_{\rm p}}{\omega^2\!+\!\Gamma_{\rm p}^2},
\end{align}
where 
$\wti\Delta_{\rm p} =\sqrt{\Delta^2_{\rm p}- \Gamma^2_{\rm p}/2}\approx\Delta_{\rm p}$,
 $\sigma_{\rm p}=
 \wti\Delta_{\rm p}\Big(\sqrt{1+
  \Gamma_{\rm p}/\wti\Delta_{\rm p} }
 -\sqrt{1-\Gamma_{\rm p}/\wti\Delta_{\rm p}}\,\Big)/2
 \approx\wti\Delta_{\rm p}/2$,
  and $\sigma_{0\rm p} 
 \approx\sqrt{1+\sqrt{2}}\Delta_{\rm p}$,
for considering the weak interaction 
between the QD and the detector with $\Gamma_{\rm p}\ll \Delta_{\rm p}$.
Since the functions of $g(\Delta_{\rm p})$
and $f(\Delta_{\rm p})$ are complex and are not explicitly displayed.
%

Next, to analyze measurement visibility, 
we define the signal as the position difference between
 the two peaks of the even and odd parities,
 ${\cal S}\equiv\wti\delta{\Delta}=|\wti\Delta_+-\wti\Delta_-|\approx\delta{\Delta}$.
Meanwhile, the noise is defined as the sum of the peak widths of two parities, 
   ${\cal N}\equiv\sigma_{+}+ \sigma_{-}\approx\frac{1}{2}(\Gamma_{+}+ \Gamma_{-})$.
 %
Subsequently, the signal-to-noise ratio (SNR) of the two characteristic peaks 
can be calculated as
\begin{align}\label{snr}
 {\text{ SNR}}&=\frac{\cal S}{ \cal N}
\approx  \frac{2\delta\Delta}{ \Gamma_{+}+ \Gamma_{-}}.
\end{align} 
This indicates that the SNR is proportional to $\delta\Delta$, which is determined by 
the intrinsic quantities involved in the system's Hamiltonian \Eq{Hs}, such as 
 coupling coefficients $\lambda_i$ and the flux $\varphi$. 
Whereas, SNR is inversely proportional to the dephasing rates ($\Gamma_{\pm}$),
 due to the back-action of the PC detector on the QD-MZM system.
Hence, SNR is influenced by both the applied
  bias voltage and temperature of the detector.

     \begin{figure}
\centering
\includegraphics[width=0.4\textwidth,clip=true]{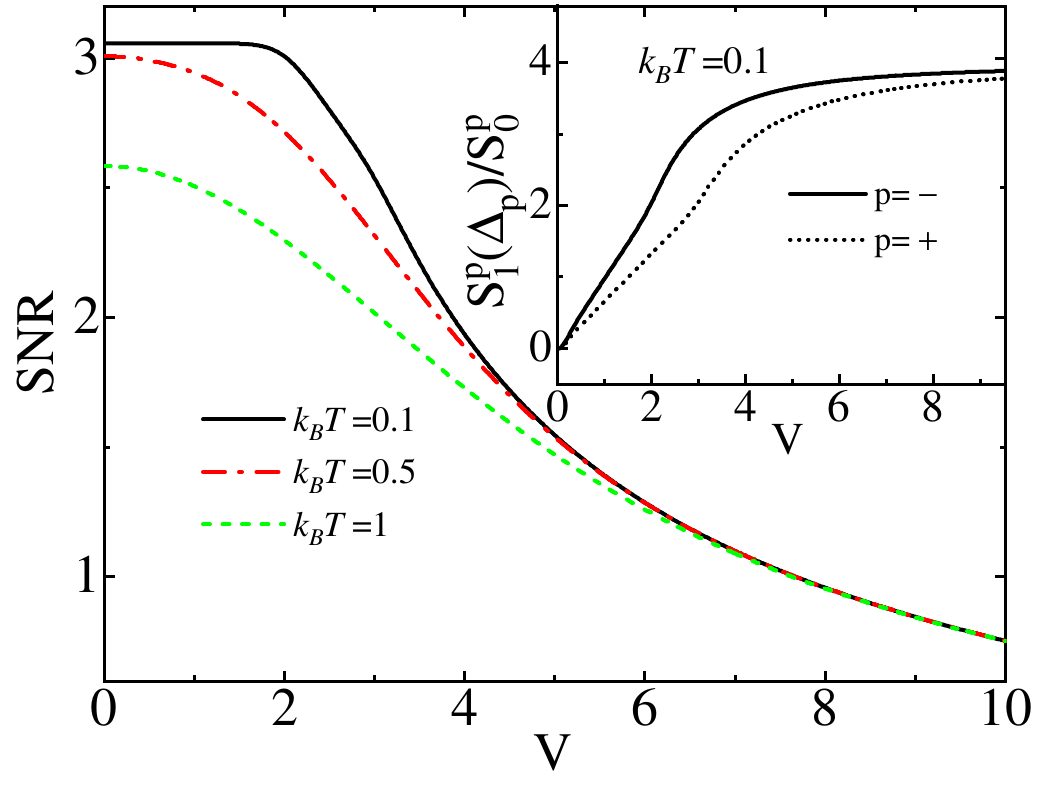}
 \caption{ (Color online) The SNR as a function 
 of the bias voltage
 with different temperature (in $\lambda$). 
 The inset is the peak-to-pedestal ratio as a function of the bias voltage
 at low temperature $k_BT=0.1\lambda$.
  The other parameters are the same as in \Fig{fig3}.
  \label{fig4}}
\end{figure}
  
  Figure \ref{fig4} depicts that the SNR 
 decreases as the measurement voltage increases, even dropping below 1 
 at high voltage limits. Furthermore, in the low bias voltage regime,
   SNR improves as the temperature decreases.
Thus, we focus on the low temperature limit with $k_BT\ll\lambda$.
Then, the SNR of \Eq{snr} simplifies to
\bsube\label{snrvol}
 \begin{align}
  {\text{ SNR}} &= \frac{4 \delta\Delta}{\eta\chi^2(\Delta_++\Delta_-)},~{\rm for}~~ V<\Delta_{\rm p},
  \label{snrlv}
  \\
{\text{ SNR}}&=\frac{2 \delta\Delta}{\eta\chi^2 V},~~~~~~~~~~~~~~{\rm for}~~  V\geq\Delta_{\rm p}. 
 \label{snrhv} 
  \end{align}
 \esube
 Seemingly, the measurement voltage should be low with
 $V<\Delta_{\rm p}$, as the 
  SNR determined by \Eq{snrlv} reaches a maximum saturation value.
  This saturation value is independent of the bias voltage
  as displayed in \Fig{fig4} at low temperature with $k_BT=0.1\lambda$.

However, from the perspective of measurement visibility,
  the height of single peaks
 in experiments must be considered. 
 Thus, we have to introduce another type of ``signal-to-noise" ratio, 
 known as {\it peak-to-pedestal}
 ratio of each characteristic peak, 
 \be
 \gamma_{\rm p}\equiv S^{\rm p}_1(\Delta_{\rm p})/S^{\rm p}_0(\Delta_{\rm p}). 
 \ee
In contrast to the SNR features discussed above, 
the peak-to-pedestal increases with the bias voltage, as shown in the inset of \Fig{fig4}.
It has the fundamental upper
bound limit of 4 at high-voltage regimes and decreases with rising temperature.
This behavior is consistent with findings in 
previous studies \cite{Kor01085312,Rus03075303,Li05066803}.

Therefore, considering the merits of SNR and peak-to-pedestal ratio,
 the optimal condition of the measurement visibility at low temperature limits
 must simultaneously satisfy both $\text{ SNR}>1$ and $\gamma_{\rm p} \rightarrow4$.
For $V<\Delta_{\rm p}$, the SNR expressed in \Eq{snrlv} readily meets
the condition of $\text{ SNR}>1$;
however, it results in $\gamma_{\rm p}<2$.
Thus, the low bias voltage regime where $V<\Delta_{\rm p}$ is not conducive to MBQ measurement.
In contrast, for $V\geq\Delta_{\rm p}$ with 
 $\gamma_{\rm p}\lesssim 4$,
 the condition of $\text{SNR}>1$
 leads to $V<\frac{2 \delta\Delta}{\eta\chi^2 }$ 
 as inferred from \Eq{snrhv}. 
Consequently, we get the optimal voltage range for MBQ readout as,
\begin{align}\label{optv}
 \Delta_{\rm p}\leq V<\frac{2 \delta\Delta}{\eta\chi^2 }.
\end{align}
 This represents another main result of this paper.
Based on \Eq{optv}, all parameters involved in the experimental measurement can be 
 appropriately selected. 
 We believe this significantly simplifies the current experimental demonstrations.

\section{Summary}
\label{thsum}

In summary,
we have thoroughly investigated the 
detector current and its power spectrum 
at arbritrary bias voltages
for the hybrid QD-MBQ system
continuously measured by the PC detector.
The analytical and numerical results have been carried out based on 
 the standard Born-Markovian quantum master equation approach. It
 maintains the detail-balance relation
 and is applicable to any bias voltage.

Our study showed that 
the readout of the MBQ from
the parity-dependent detector current ( $I_{\rm p}$) or the dot occupation number ($n_{\rm p}$)
is only suitable for specific conditions:
the off-resonance of dot ($\varepsilon_{\rm d}\neq 0$)
and under low bias voltage ($V\lesssim\Delta_{\rm p}$).
The former is because
 the coherent dynamics in the two parity spaces
 exhibit the equivalent Rabi resonances at $\varepsilon_{\rm d}= 0$.
The latter is due to 
the fact that the parity information in
 $I_{\rm p}$ or $n_{\rm p}$ 
relies on the energy-dependent bath correlation function ($C^{(\pm)}(\pm \Delta_{\rm p})$),
which accounts for the energy exchange ($\Delta_{\rm p}$) between
the measured system and detector. 
Nevertheless, at large bias voltage regime ($V\gg \Delta_{\rm p}$),
   $\Delta_{\rm p}$ becomes  
  negligible in the bath correlation function, rendering
  $I_{\rm p}$ or $n_{\rm p}$  
insensitive to the parity.

In contrast, the parity-dependent current power spectrum $S_{\rm p}(\omega)$, 
allows for the MBQ readout 
  from the parity-dependent
 Rabi oscillation peak signals across all bias voltages,
  without restrictions 
 on the dot energy level.
This is because the current power spectrum
 not only captures the back action of the detector through
the bath correlation, but also reflects the intrinsic coherent dynamics
of the measured system, 
characterized by its parity-dependent Rabi frequencies ($\Delta_{\rm p}$).
Particularly, for the measurement visibility
 analysis, we evaluate the signal-to-noise ratio (SNR) of the
two parity-dependent characteristic peaks and 
  the peak-to-pedestal ratio for each individual peak.
 These metrics enabled us to determine  
  the optimal bias voltage range for the PC detector 
 to readout the MBQ 
 under low temperature conditions.
The results of this study provide
significant insights for current experimental demonstrations.

\acknowledgments
We acknowledge helpful discussions with
 Prof. Xin-Qi Li and 
 Prof. YiJing Yan.
  The support from the National Natural Science Foundation of China
(Grant No. 12175052) is acknowledged.


\end{document}